\documentclass[a4paper,11pt]{article}
\pdfoutput=1
\usepackage{jcappub} 
\usepackage{amsmath}
\usepackage{amssymb}
\usepackage{dcolumn}
\usepackage{graphicx}
\usepackage{verbatim}
\usepackage{color}

\newcommand{\be}{\begin{equation}}
\newcommand{\ee}{\end{equation}}
\newcommand{\f}{\frac}
\newcommand{\M}{\mathcal M}
\newcommand{\ep}{\epsilon}

\newcommand{\ph}{\phi_{\rm VEV}}
\newcommand{\Mpl}{M_{\mathrm{Pl}}}
\newcommand{\A}{A_{\rm VEV}}
\newcommand{\V}{V_{\rm VEV}}

\begin{document}
\title{Angular momentum loss for eccentric compact binary in screened modified gravity}
\author[a,b]{Xing Zhang}
\author[a,b]{Wen Zhao}
\author[a,b]{Tan Liu}
\author[c,d]{Kai Lin}
\author[e]{Chao Zhang}
\author[e]{Xiang Zhao}
\author[f]{Shaojun Zhang}
\author[f]{Tao Zhu}
\author[e,f]{Anzhong Wang}
\affiliation[a]{CAS Key Laboratory for Researches in Galaxies and Cosmology, Department of Astronomy, \\ University of Science and Technology of China, Chinese Academy of Sciences, Hefei, Anhui 230026, China}
\affiliation[b]{School of Astronomy and Space Science, University of Science and Technology of China, Hefei 230026, China}
\affiliation[c]{Hubei Subsurface Multi-scale Imaging Key Laboratory, Institute of Geophysics and Geomatics, China University of Geosciences, Wuhan, Hubei, 430074, China}
\affiliation[d]{Escola de Engenharia de Lorena, Universidade de S\~ao Paulo, 12602-810, Lorena, SP, Brazil}
\affiliation[e]{GCAP-CASPER, Physics Department, Baylor University, Waco, TX 76798-7316, USA}
\affiliation[f]{Institute for Advanced Physics $\&$ Mathematics, Zhejiang University of Technology, Hangzhou 310032, China}
\emailAdd{starzhx@ustc.edu.cn}
\emailAdd{wzhao7@ustc.edu.cn}

\abstract{Gravitational wave (GW) observations provide insight into the gravity regime. These observations relies on the comparison of the data to the GWs model, which depends on the orbital evolution of the binary systems. In this paper, we study the orbital evolution of eccentric compact binaries within screened modified gravity, which is a kind of scalar-tensor theory with screening mechanisms. According to the Noether's theorem, from the theory's GW action, we compute the angular momentum flux both as a function of the fields in the theory and as a function of the multipole moments of a N-body system. We specialize to the orbital parameters of eccentric compact binaries to express its orbital energy and angular momentum decay, and then derive the decay rates of its orbital eccentricity and semimajor axis. We find that these quantities decay faster than in General Relativity due to the presence of dipole radiation.}

\keywords{modified gravity, binary pulsar, gravitational waves, angular momentum loss}

\maketitle
\flushbottom

\section{Introduction}\label{section1}
General Relativity (GR) has been very successful from laboratory tests \cite{Hoyle:2001ab,Adelberger:2001aa} to solar system \cite{Will:1993aa,Will:2014aa} and binary pulsar \cite{Stairs:2003aa,Manchester:2015aa,Kramer:2016aa,Wex:2014aa} tests. Nevertheless, testing GR still is one of the key tasks for modern physics. As well known, GR is incomplete in the ultraviolet regime where it should be replaced by a still unknown quantum gravity \cite{Kiefer:2007aa,DeWitt:1967yk}. In addition, the evolution of the Universe cannot be properly described by GR unless one adds the dark matter and dark energy in our Universe \cite{Cline:2013aa,Sahni:2004ai}. As the minimal extension of GR, scalar-tensor gravity has been widely investigated in comparison with GR in the literature \cite{Fujii:2003aa,Faraoni:2004pi,Damour:1992ab}, which invokes a conformal coupling between matter and scalar fields. In order to suppress the scalar force (i.e., fifth force \cite{Gubser:2004aa,Adelberger:2009aa,Williams:2012aa}) and allow gravity theories to evade the tight gravitational tests in the solar system and the laboratory, we need the screening mechanisms which operate environmentally dependent scalar fields. The screening mechanisms in scalar-tensor gravity mainly include the chameleon \cite{Khoury:2004aa,Khoury:2004ab}, symmetron \cite{Hinterbichler:2010es,Hinterbichler:2011aa}, dilaton \cite{Damour:1994ab,Brax:2010aa} and $f(R)$ \cite{Sotiriou:2010aa,De-Felice:2010aa} theories.  These theories can be described within a unified theoretical framework called screened modified gravity (SMG) \cite{Brax:2012aa}.

In order that SMG can generate the screening mechanisms, the effective potential of the scalar field must have a physical vacuum \cite{Brax:2012aa}. Besides, the effective mass of the scalar field around this vacuum is required to increase as the ambient density increases. Therefore, the scalar field can be screened and evade the tests in high density regions (e.g., the solar system), where the range of the fifth force is so short that it cannot be detected within current experimental accuracy \cite{Khoury:2004ab,Gubser:2004aa}. Whereas in low density regions (e.g., the Galaxy and the Universe), the long-range fifth force may affect galactic dynamics \cite{Gronke:2015aa,Schmidt:2010aa}, and the scalar field can play the role of dark energy to accelerate the expansion of the Universe \cite{Khoury:2004aa,Hinterbichler:2011aa}. In addition, the tensor sector of SMG, like in GR, the tensor gravitational waves (GWs) contain two basic polarization states and propagate with the speed of light, which satisfy the constraints of GWs speed from GW170817 \cite{Abbott:2017ac,Abbott:2017aa}.

The GWs observations provide an excellent opportunity to look into the extreme gravity regime of coalescing binary, where gravity is strong, dynamical and non-linear. The interpretation of GWs observations relies on the comparison of the data to the GWs model, which depends on the orbital evolution of the binary systems, and in particular on its orbital energy and angular momentum decay. In previous work \cite{Zhang:2017aa}, we calculated the GWs energy flux in terms of derivatives of multipole moments in SMG, and derived the rate of the orbital energy loss for compact binaries in quasi-circular orbits. Based on it, in the literature \cite{Liu:2018ab}, we calculated the waveforms of GWs emitted by inspiralling compact binaries. Recently, the eccentric GWs have been widely studied in the literature \cite{Loutrel:2017aa,Ma:2017aa}. However, in our previous works, we considered only the quasi-circular orbits. As an extension of this issue, in this paper we study the decay rates of the orbital energy and angular momentum for compact binaries in quasi-elliptical orbits. We use the Noether's theorem to compute the angular momentum flux carried by all propagating modes in terms of derivatives of these fields, and use the multipolar expansion of the fields to calculate this flux in terms of derivatives of multipole moments. We use a Keplerian parametrization to present the energy and angular momentum fluxes as the functions of the orbital parameters of eccentric compact binary systems. We conclude by combining this with the orbital energy decay to find expressions for the decay rates of the orbital eccentricity and semimajor axis of this system. We find that generically these decay rates are typically faster in SMG than in GR due to the presence of dipole radiation.

This paper is organized as follows. In Sec. \ref{section2}, we display the action for SMG and derive the field equations and their weak-field limit. In Sec. \ref{section3}, we calculate the angular momentum loss rate due to the GWs emission in SMG. In Sec. \ref{section4}, we derive the decay rates of the orbital eccentricity and semimajor axis. We conclude in Sec. \ref{section5} with a summary and discussion.

Throughout this paper, the metric convention is chosen as $(-,+,+,+)$, Greek indices ($\mu, \nu, \cdots$) run over $0, 1, 2, 3$, and Latin indices ($i, j, \cdots$) run over $1, 2, 3$. We set the units to $c=\hbar=1$, and therefore the reduced Planck mass is $M_\text{Pl} = \sqrt{1/8 \pi G}$, where $G$ is the gravitational constant.

\section{Screened Modified Gravity}\label{section2}
\subsection{Action}
SMG is a kind of scalar-tensor theory with screening mechanisms, which can suppress the fifth force in dense regions and pass the solar system tests \cite{Zhang:2016aa}. A general scalar-tensor gravity with two arbitrary functions is given by the following action \cite{Damour:1992ab,Brax:2012aa}:
\begin{align} \label{action_0}
\begin{split}
S=&\int d^4x\sqrt{-g}\left[\frac{\Mpl^2}{2}R-\frac12(\nabla\phi)^2-V(\phi)\right]+S_m\left[A^2(\phi) g_{\mu\nu},\,\psi_m^{(i)}\right],
\end{split}
\end{align}
where $g$ is the determinant of the metric $g_{\mu\nu}$, $R$ is the Ricci scalar, $\psi_m^{(i)}$ are various matter fields labeled by $i$\,. The potential $V(\phi)$ characterizes the scalar self-interaction, which has three main effects in the theory: First, it can play the role of dark energy. Second, it endows the scalar field with mass. Finally, it may introduce nonlinearities into the scalar dynamics. $A(\phi)$ is a conformal coupling function characterizing the interaction between scalar field $\phi$ and matter fields $\psi_m$, which induces the fifth force in the theory.

In general, the scalar field is governed by the effective potential $V_{\text {eff}}(\phi)$ in Eqs. \eqref{scalar_eom} and \eqref{Veff2}. In order to generate a screening effect to suppress the fifth force in high density environments, the effective potential of the scalar field must have a minimum \cite{Brax:2012aa,Zhang:2016aa},
\begin{align}\label{mass_eff}
\frac{\rm d V_{\rm eff}}{\rm d\phi}\bigg|_{\phi_{\rm min}}=0\,,\quad~    m^2_{\rm eff}\equiv \frac{\rm d^2 V_{\rm eff}}{\rm d\phi^2}\bigg|_{\phi_{\rm min}}>0\,,
\end{align}
which can be naturally understood as a physical vacuum. Around this physical vacuum, the scalar field acquires an effective mass, which is required to increase as the ambient density increases, i.e., ${\rm d} m_{\rm eff}(\rho)/{\rm d}\rho>0$. Therefore, the scalar force can be suppressed in high density regions, where the force range is so short that it cannot be detected. This kind of scalar-tensor gravity with screening mechanism is often called SMG \cite{Brax:2012aa,Brax:2014aa}, which can generate the screening effect to suppress the fifth force in high density environments and pass the solar system and laboratory tests.

\subsection{Matter action and field equations}
In general, so long as the compact objects are far enough from each other, their motion can be effectively described through point particles with the composition-dependent effects encapsulated in nonstandard couplings in the particle action. Eardley \cite{Eardley:1975aa} first showed that these effects could be accounted for by the matter action for a system of point-like masses,
\begin{align}
\label{matter_action}
\begin{split}
S_m=&-\sum_a\int m_a(\phi)d\tau_a\,,
\end{split}
\end{align}
where $m_a(\phi)$ is the $\phi$-dependent mass of the $a$-th point-particle, and $\tau_a$ is its proper time measured along its worldline $x^\lambda_a$.

Now, the full action for a system of point-like masses is
\begin{align} \label{action_1}
\begin{split}
S=&\int d^4x\sqrt{-g}\left[\frac{\Mpl^2}{2}R-\frac12(\nabla\phi)^2-V(\phi)\right]-\sum_a\int m_a(\phi)d\tau_a\,.
\end{split}
\end{align}
The variation of this action with respect to the tensor field and the scalar field yields the tensor field equation
\begin{align}
G_{\mu\nu}\label{tensor_eom}
= 8 \pi G \left[T_{\mu\nu}(x,\phi)+T_{\phi\mu\nu}(\phi)\right]\,,
\end{align}
and the scalar field equation
\begin{align}
\square_g\phi\label{scalar_eom}
=\frac{\partial V_{\rm eff}(\phi)}{{\partial }\phi}\,,
\end{align}
where $G_{\mu\nu}$ is the Einstein tensor, $T^{\mu\nu}(x,\phi)\equiv(2/\sqrt{-g})\delta S_m/\delta g_{\mu\nu}$ is the matter energy-momentum tensor, and $\square_g\equiv(-g)^{-1/2}\partial_{\nu}((-g)^{1/2}g^{\mu\nu}\partial_{\mu})$ is the curved space d'Alembertian. $T_{\phi\mu\nu}(\phi)$ is the  energy-momentum tensor of the scalar field, given by
\begin{align}\label{Tuv_phi}
T_{\phi\mu\nu}(\phi)=\partial_\mu\phi\partial_\nu\phi-g_{\mu\nu}\left[\frac{1}{2}(\partial\phi)^2+V(\phi)\right].
\end{align}
For a negligibly self-gravitating body the effective potential is
\be\label{Veff2}
V_{\text {eff}}(\phi) = V(\phi)+{\rho}A(\phi)\,,
\ee
where $\rho$ is the matter density of the local environment of the scalar field.

\subsection{Weak-field limit}
In the weak-field limit, the tensor field $g_{\mu\nu}$ and the scalar field $\phi$ can be expanded around the backgrounds as follows:
\begin{align}\label{perturbations}
g_{\mu\nu}=\eta_{\mu\nu}+h_{\mu\nu}\,,\qquad \phi=\ph+\varphi\,\,,
\end{align}
where $\eta_{\mu\nu}$ is the flat Minkowski background, and $\ph$ is the vacuum expectation value (VEV) of the scalar field (i.e., scalar background).

The bare potential $V(\phi)$ and the coupling function $A(\phi)$ can be expanded in Taylor's series around the scalar background as follows:
\begin{align}
V(\phi)=\V+V_1\varphi+V_2\varphi^2+\mathcal{O}(\varphi^3),
\end{align}
\begin{align}
A(\phi)=\A+A_1\varphi+A_2\varphi^2+\mathcal{O}(\varphi^3),
\end{align}
where $\A\equiv A(\ph)$ is the coupling function VEV, and $\V\equiv V(\ph)$ is the bare potential VEV which can act as the cosmological constant to accelerate the expansion of the late Universe \cite{Zhang:2016aa}.

The mass $m_a(\phi)$ for a strongly self-gravitating body can be expanded in Taylor's series around the scalar background,
\begin{align}
\label{inertial_mass}
\begin{split}
m_a(\phi)=&m_a\bigg[1+s_a\Big(\frac{\varphi}{\phi_{\rm VEV}}\Big)+\mathcal{O}\Big(\frac{\varphi}{\phi_{\rm VEV}}\Big)^2\bigg]\,,
\end{split}
\end{align}
where $m_a\equiv m_a(\ph)$ is the inertial mass at the scalar background, and the sensitivity $s_a$ is defined by \cite{Alsing:2012aa}
\begin{align}
\label{sensitivities}
s_a\equiv\frac{\partial(\ln m_a)}{\partial(\ln \phi)}\bigg|_{\phi_{\rm VEV}},
\end{align}
which connects with the scalar charge (i.e., screened parameter) $\ep_a$ by \cite{Zhang:2017aa}
\begin{align}\label{s_a}
s_a=\frac{\ph}{2\Mpl}\ep_a,
\end{align}
with
\begin{align}\label{epsilon_a}
\ep_a\equiv\frac{\ph-\phi_a}{M_\text{Pl}\Phi_{a}}\,,
\end{align}
where $\Phi_a=Gm_a/R_a$ is the compactness (i.e., negative Newtonian gravitational potential at the surface) of the $a$-th object, and $\phi_a$ is the position of the minimum of $V_{\rm eff}$ inside the $a$-th object.

In the weak-field limit, imposing the Lorentz gauge condition $\partial^{\mu}(h_{\mu\nu}-\frac{1}{2}\eta_{\mu\nu}h)=0$, the field equations \eqref{tensor_eom} and \eqref{scalar_eom} reduce to
\begin{align}\label{linear_tensor_eq}
\square{h}_{\mu\nu}=-16\pi G\Big(T_{\mu\nu}-\f{1}{2}\eta_{\mu\nu}T\Big),
\end{align}
\begin{align}\label{linear_scalar_eq}
(\square-m^2_s)\varphi=-\frac{\partial T}{\partial\varphi}+h^{\mu\nu}\partial_{\mu}\partial_{\nu}\varphi,
\end{align}
where $\square\equiv\eta^{\mu\nu}\partial_{\mu}\partial_{\nu}$ is the flat-space d'Alembertian, and
\begin{align}
m^2_{s}\equiv \frac{\rm d^2 V_{\rm eff}}{\rm d\phi^2}\bigg|_{\ph}\!\!=2(V_2+\rho_b A_2)\,
\end{align}
is the effective mass of the scalar field in a homogeneous background density $\rho_b$.

\section{Angular Momentum Loss Rate}\label{section3}
In this section, we calculate the angular momentum loss rate due to the emission of tensor and scalar gravitational radiations in SMG. We begin with a generic calculation by using the Noether's theorem. We then specialize these results to a binary system and obtain the angular momentum loss rate in this system.

\subsection{General calculation}
 Now, let us compute the angular momentum of GWs in SMG. Angular momentum is the conserved charge associated to invariance under spatial rotations. In the weak-field limit, from the action \eqref{action_1}, the Lagrangians of the tensor and scalar GWs reduce to
\begin{align}
\mathcal{L}_T=-\f{\Mpl^2}{8}\partial_{\mu} h_{ij}^{\rm TT}\partial^{\mu} h_{ij}^{\rm TT}
\end{align}
\begin{align}
\mathcal{L}_{S}=-\f{(\partial\varphi)^2}{2}-\f{1}{2}m^2_s\varphi^2,
\end{align}
where ${h}^{\rm TT}_{ij}$ is the transverse-traceless (TT) part of ${h}_{ij}$.
The angular momentum carried by GWs can be derived directly from these Lagrangians by investigating the Noether charges and currents in the theory. We then obtain the angular momentum fluxes of the tensor and scalar GWs,
\begin{align}\label{tensor_ang_moment}
\dot{L}^i_{T}=-\epsilon^{ijk}\f{r^2}{32\pi G}\int{d\Omega}\big\langle2h_{jl}^{\rm TT}\dot{h}_{kl}^{\rm TT}-\dot{h}_{lm}^{\rm TT}x^j\partial_kh_{lm}^{\rm TT}\big\rangle
\end{align}
\begin{align}\label{scalar_ang_moment}
\dot{L}^i_{S}=\epsilon^{ijk}r^2\int{d\Omega}\big\langle\dot{\varphi} x^j\partial_k\varphi\big\rangle,
\end{align}
where $\epsilon^{ijk}$ is the Levi-Civita symbol, $\Omega$ is the solid angle, the overdots denote derivatives with respect to coordinate time, and the angular brackets represent a time average over a period of the system's motion. The tensor angular momentum flux is exactly the same as that in GR. The first term in Eq. \eqref{tensor_ang_moment} comes from the spin angular momentum of the tensor graviton, while the second is the contribution from the orbital angular momentum of the tensor GWs. In Eq. \eqref{scalar_ang_moment}, the scalar angular momentum flux comes only from the orbital angular momentum of the scalar GWs, because the scalar field is spin-0.

\subsection{Angular momentum flux for N-body systems}
The solutions of the field equations \eqref{linear_tensor_eq} and \eqref{linear_scalar_eq} for N-body systems were calculated in our previous work \cite{Zhang:2017aa},
\begin{align}\label{h_ij_M}
h_{ij}^{\rm TT}(t,\mathbf r)=\f{2G}{r}\ddot{M}^{\rm TT}_{ij}(t-r),
\end{align}
\begin{align}\label{t_dep_sca_sol3}
\begin{split}
\varphi(t,\mathbf{r})=&-\Mpl\f{G}{r}\sum_{{\ell}=0}^\infty\f{1}{\ell!}n_{i_1}n_{i_2}\!\cdots\! n_{i_\ell}{\partial^{\ell}_t}\M_{\ell}^{i_1i_2\cdots i_{\ell}}\bigg|_{\rm ret}\,,
\end{split}
\end{align}
where the superscript `TT' stands for the transverse-traceless part, the subscript `ret' means that the quantity $M^{ij}$ is evaluated at the retarded time $t-r$, and $\mathbf{n}=\mathbf{r}/r$ is the unit vector in the $\mathbf{r}$ direction. We have defined the multipole moments for a N-body system,
\begin{align}\label{mass_quadrupole}
M^{ij}(t)=\sum_a m_a r_a^i(t) r_a^j(t)\,,
\end{align}
\begin{align}\label{mass_moment_0}
\begin{split}
\M_{\ell}^{i_1i_2\cdots i_{\ell}}(t)=&\sum_a\ep_aM_a(t)r_a^{i_1}(t)r_a^{i_2}(t)\cdots r_a^{i_{\ell}}(t),
\end{split}
\end{align}
and the mass
\begin{align}\label{t_dep_mass}
\begin{split}
M_a(t)=& m_a\Big[1-\f12v_a^2(t)-\sum_{b\ne a}\f{Gm_b}{r_{ab}(t)}\Big]\,,
\end{split}
\end{align}
where we have neglected the scalar field mass, which roughly is cosmological scales.

With this at hand, from Eqs. \eqref{tensor_ang_moment} and \eqref{scalar_ang_moment}, we obtain the angular momentum fluxes of the tensor and scalar GWs,
\begin{align}\label{tensor_ang_moment1}
\dot{L}^{i}_{T}=-\f{2G}{5}\epsilon^{ijk}\left\langle{\ddot{M}^{jl}\dddot{M}^{kl}}\right\rangle,
\end{align}
\begin{align}\label{scalar_ang_moment1}
\dot{L}^i_{S}=-\epsilon^{ijk}\f{G}{6}\Big\langle\dot\M^j_1\ddot\M^k_1+\f{1}{5}\ddot\M^{jl}_2\dddot\M^{kl}_2+\f{1}{5}\dddot\M^{llj}_3\ddot\M^k_1\Big\rangle\,.
\end{align}
In Eq. \eqref{scalar_ang_moment1}, the first and second terms are the contributions of the dipole and quadrupole radiations, and the last term represents the dipole-octupole cross term. Because of spin-0 scalar field, the scalar monopole radiation carries energy but not angular momentum.

\subsection{Specialization to binary systems}
Now let us consider a binary pulsar system with quasi-elliptical orbit, which is parameterized in the center of mass frame by
\begin{align}\label{x_y_z}
\begin{split}
x_n(t)&=-r_n\cos\theta,~~~y_n(t)=-r_n\sin\theta,~~~z_n=0,
\end{split}
\end{align}
with
\begin{align}\label{rn12}
\begin{split}
r_n=\f{a_n(1-e^2)}{1-e\cos{\theta}},
\end{split}
\end{align}
where $e$ and $\theta$ are the orbital eccentricity and true anomaly,  $a_n$ is the semimajor axis, and $n=1, 2$ denote the pulsar and its companion, respectively. Using these, from Eqs. \eqref{tensor_ang_moment1} and \eqref{scalar_ang_moment1}, we obtain the angular momentum decay rates for multipole radiations as follows:
\begin{align}\label{L_tq}
\dot{L}_{T}^{q}&=-\f{32}{5}\f{G\mu^2{(Gm)}^{5/2}}{a^{7/2}}\f{1}{(1-e^2)^2}\Big(1+\f{7}{8}e^2\Big)\Big(1+\f{5}{4}\ep_1\ep_2\Big),
\end{align}
\begin{align}\label{L_sd}
\begin{split}
\dot{L}_{S}^d=&-\f{G\mu^2(Gm)^{3/2}}{6a^{5/2}}\f{\epsilon_d^2}{(1-e^2)}+\f{G\mu^2(Gm)^{5/2}}{6a^{7/2}}
\f{\epsilon_d}{(1-e^2)^2}\big[2(\epsilon_{d1}+\epsilon_{d2})+e^2(\epsilon_{d1}+4\epsilon_{d2})\big],
\end{split}
\end{align}
\begin{align}\label{L_sq}
\dot{L}_{S}^q=-\f{8}{15}\f{G\mu^2(Gm)^{5/2}}{a^{7/2}}\epsilon_q^2\f{(1+\f{7}{8}e^2)}{(1-e^2)^2},
\end{align}
\begin{align}\label{L_sdo}
\dot{L}_{S}^{do}=\f{G\mu^2(Gm)^{5/2}}{30a^{7/2}}\epsilon_d\epsilon_o\f{(1-\f{17}{2}e^2)}{(1-e^2)^2},
\end{align}
where $m$ and $\mu$ are the total and reduced masses of the system, and  $d$, $q$, and $o$ denote dipole, quadrupole, and octupole, respectively. We have defined
\begin{subequations}\label{epsilon_i}
\begin{align}
\ep_d\equiv\ep_2-\ep_1,
\end{align}
\begin{align}
\ep_{d1}\equiv\f{1}{m}(\ep_2m_1-\ep_1m_2),
\end{align}
\begin{align}
\ep_{d2}\equiv\f{1}{2m^2}(\ep_2m_1^2-\ep_1m_2^2),
\end{align}
\begin{align}
\ep_q\equiv\f{1}{m}(\ep_2m_1+\ep_1m_2),
\end{align}
\begin{align}
\ep_{o}\equiv\f{1}{m^2}(\ep_2m_1^2-\ep_1m_2^2)=2\ep_{d2},
\end{align}
\end{subequations}
where $\ep_1$ and $\ep_2$ are the scalar charges of the pulsar and its companion, and defined in Eq. \eqref{epsilon_a}. By summing Eqs. \eqref{L_tq}-\eqref{L_sdo}, we obtain the total angular momentum decay rate as follows:
%
\begin{align}
\begin{split}\label{total_ang_mom_flux}
\dot{L}=&-\f{G\mu^2(Gm)^{3/2}}{6a^{5/2}}\f{\epsilon_d^2}{(1-e^2)}-\f{G\mu^2(Gm)^{5/2}}{a^{7/2}}\f{1}{(1-e^2)^2}\bigg\{\f{32}{5}\big(1+\f{7}{8}e^2\big)\big(1+\f{5}{4}\ep_1\ep_2\big)
\\&-\f{1}{6}\Big[2\big(\epsilon_{d1}+\epsilon_{d2}\big)+e^2\big(\epsilon_{d1}+4\epsilon_{d2}\big)\Big]\epsilon_d+\f{1}{15}\big(8+7e^2\big)\epsilon_q^2-\f{1}{60}\big(2-17e^2\big)\epsilon_d\epsilon_o\bigg\}.
\end{split}
\end{align}
%
It can be seen that the first term is the dipole radiation, which dominates the angular momentum decay. In the limit of $\ep_1$ and $\ep_2$ $\rightarrow0$, the above expression reduces to the GR results.

\subsection{Energy flux}
In previous work \cite{Zhang:2017aa}, we have calculated the rate of the orbital energy loss in SMG for compact binaries in quasi-circular orbits. As an extension, in this subsection we consider the case of quasi-elliptical orbits.

According to our work \cite{Zhang:2017aa}, we obtain the tensor and scalar energy fluxes in terms of derivatives of multipole moments as follows:
\begin{align}\label{Q_F_g}
\begin{split}
\dot{E}_T=-\f{G}{5}\Big\langle\dddot{M}^{kl}\dddot{M}^{kl}-\f{1}{3}\big(\dddot{M}^{kk}\big)^2\Big\rangle,
\end{split}
\end{align}
\begin{align}
\label{scalar_flux}
\begin{split}
\dot{E}_{S}=&-\frac{G}{2}\Big{\langle}\dot{\M}_0\dot{\M}_0+\frac{1}{3}\ddot{\M}_1^k\ddot{\M}_1^k+\frac{1}{3}\dot{\M}_0\dddot{\M}_2^{kk}
+\frac{1}{30}\dddot{\M}_2^{kl}\dddot{\M}_2^{kl}+\frac{1}{60}\dddot{\M}_2^{kk}\dddot{\M}_2^{ll}+\frac{1}{15}\ddot{\M}_1^{k}\ddddot{\M}_3^{kll}\Big{\rangle},
\end{split}
\end{align}
where the angular brackets represent a time average over a period of the system's motion, the overdots denote derivatives with respect to coordinate time, $M^{ij}$ and $\M_{\ell}^{i_1i_2\cdots i_{\ell}}$ are the multipole moments defined in Eqs. \eqref{mass_quadrupole} and \eqref{mass_moment_0}, and we have neglected the scalar field mass of cosmological scales. In Eq. \eqref{scalar_flux}, the first and second terms are the monopole and dipole radiations, the third term represents the monopole-quadrupole cross term, the fourth and fifth terms are the quadrupole radiations, and the last term represents the dipole-octupole cross term.

Using Eqs. \eqref{x_y_z} and \eqref{rn12}, from these expressions we obtain the orbital energy decay rates caused by multipole radiations for eccentric compact binaries as follows:
\begin{align}
\dot{E}_{T}^{q}&=-\f{32}{5}\f{G\mu^2{(Gm)}^3}{a^5}\f{\big(1+\f{73}{24}e^2+\f{37}{96}e^4\big)}{(1-e^2)^{7/2}}\big(1+\f{3}{2}\ep_1\ep_2\big),
\end{align}
\begin{align}
\dot{E}_{S}^m=-\f{G^4\mu^2m^3}{4a^5}\ep^2_m\f{e^2\big(1+\f{e^2}{4}\big)}{(1-e^2)^{7/2}},
\end{align}
\begin{align}
\begin{split}
\dot{E}_{S}^d=&-\f{G^3\mu^2m^2}{6a^4}\ep^2_d\f{\big(1+\f{e^2}{2}\big)}{(1-e^2)^{5/2}}+\f{G^4\mu^2m^3}{3a^5}\f{\ep_d}{(1-e^2)^{7/2}}
\\&\times
\bigg[\ep_{d1}\!+\!\ep_{d2}+\big(3\ep_{d1}\!+\!\f{13}{2}\ep_{d2}\big)e^2+\big(\f{3}{8}\ep_{d1}\!+\!\f{5}{4}\ep_{d2}\big)e^4\bigg],
\end{split}
\end{align}
\begin{align}
\dot{E}_{S}^q=-\f{8G^4\mu^2m^3}{15a^5}\ep^2_q\f{\big(1+\f{99}{32}e^2+\f{51}{128}e^4\big)}{(1-e^2)^{7/2}},
\end{align}
\begin{align}
\dot{E}_{S}^{mq}=\f{G^4\mu^2m^3}{6a^5}\ep_m\ep_q\f{e^2\big(1+\f{e^2}{4}\big)}{(1-e^2)^{7/2}},
\end{align}
\begin{align}
\dot{E}_{S}^{do}=\f{G^4\mu^2m^3}{30a^5}\ep_d\ep_o\f{\big(1-18e^2-\f{39}{8}e^4\big)}{(1-e^2)^{7/2}},
\end{align}
where $m$, $d$, $q$, and $o$ denote monopole, dipole, quadrupole, and octupole, respectively. The quantity $\ep_m\equiv\ep_1+\ep_2+(\ep_1m_2+\ep_2m_1)/m$, and the quantities $\ep_d$, $\ep_{d1}$, $\ep_{d2}$, $\ep_q$, and $\ep_o$ are defined in Eqs. \eqref{epsilon_i}. By summing the above expressions, we obtain the total energy decay rate as follows:
%
\begin{align}
\begin{split}\label{total_energy_flux}
\dot{E}=&-\f{G^3\mu^2m^2}{6a^4}\ep^2_d\f{\big(1+\f{e^2}{2}\big)}{(1-e^2)^{5/2}}-\f{G^4\mu^2m^3}{a^5}\f{1}{(1-e^2)^{7/2}}\bigg\{\f{32}{5}\Big(1+\f{73}{24}e^2+\f{37}{96}e^4\Big)\Big(1+\f{3}{2}\ep_1\ep_2\Big)
\\&
+\f{e^2}{4}\big(1+\f{e^2}{4}\big)\ep_m^2-\ep_d\Big[\f13\big(\ep_{d1}+\ep_{d2}\big)+\big(\ep_{d1}+\f{13}{6}\ep_{d2}\big)e^2+\big(\f{1}{8}\ep_{d1}+\f{5}{12}\ep_{d2}\big)e^4\Big]
\\&
+\f{8}{15}\big(1+\f{99}{32}e^2+\f{51}{128}e^4\big)\ep^2_q-\f{e^2}{6}\big(1+\f{e^2}{4}\big)\ep_m\ep_q-\f{1}{30}\big(1-18e^2-\f{39}{8}e^4\big)\ep_d\ep_o\bigg\}\,,
\end{split}
\end{align}
%
where the first term represents the dipole radiation, and the second term is the contributions of the quadrupole/monopole and the cross terms. Similarly to the angular momentum decay, the dipole radiation dominates the energy decay. Similarly, we also find that this result reduces to that in GR, in the limit of $\epsilon_1\rightarrow 0$ and $\epsilon_2\rightarrow 0$.

\section{Orbital Dynamics}\label{section4}
Because of the conformal coupling between matter and scalar field, the scalar force modifies the conservative orbital dynamics of binary systems, which was derived in \cite{Zhang:2017aa}. At Newtonian order, the effects of the scalar force are equivalent to redefining the gravitational coupling constant between a binary system as $\mathcal G=G\left(1+\frac12\ep_1\ep_2\right)$ \cite{Zhang:2017aa}.

In a binary system, without radiation-reaction, the orbital binding energy $E$ and the orbital angular momentum $L$ are constants, and related to the orbital eccentricity $e$ and semimajor axis $a$ via,
\begin{align}
E=-\f{\mathcal Gm\mu}{2a},
\end{align}
\begin{align}
L^2={\mathcal G}m\mu^2a(1-e^2).
\end{align}
Once considering radiation-reaction, the orbital eccentricity $e$ and semimajor axis $a$ are no longer constants, and their time derivatives are related to the orbital energy and angular momentum fluxes via,
\begin{align}
\dot{a}=\f{2a^2}{\mathcal Gm\mu}\dot{E},
\end{align}
\begin{align}\label{e_t_derivative_0}
\dot{e}=\f{a(1-e^2)}{{\mathcal G}m\mu e}\Big[\dot{E}-\f{(\mathcal Gm)^{1/2}}{a^{3/2}(1-e^2)^{1/2}}\dot{L}\Big]\,,
\end{align}
where $\dot{E}<0$ is a negative contribution to $\dot{e}$, $\dot{L}<0$ is a positive contribution to $\dot{e}$, and the contribution of $\dot{E}$ is greater than that of $\dot{L}$, because $\dot{e}$ in Eq. \eqref{e_t_derivative_1} is negative.

Substituting Eqs. \eqref{total_ang_mom_flux} and \eqref{total_energy_flux} into the above expressions, we obtain the decay rates of the orbital semimajor axis and eccentricity,
%
\begin{align}
\begin{split}\label{a_t_derivative_1}
\dot{a}=&-\f{G^2\mu m}{3a^2}\ep^2_d\f{\big(1+\f{e^2}{2}\big)}{(1-e^2)^{5/2}}-\f{G^3\mu m^2}{a^3}\f{1}{(1-e^2)^{7/2}}\bigg\{\f{64}{5}\Big(1+\f{73}{24}e^2+\f{37}{96}e^4\Big)\big(1+\ep_1\ep_2\big)
\\&
+\f{e^2}{2}\big(1+\f{e^2}{4}\big)\ep_m^2-\ep_d\Big[\f{2}{3}\big(\ep_{d1}+\ep_{d2}\big)+\big(2\ep_{d1}+\f{13}{3}\ep_{d2}\big)e^2+\big(\f{1}{4}\ep_{d1}+\f{5}{6}\ep_{d2}\big)e^4\Big]
\\&
+\f{16}{15}\big(1+\f{99}{32}e^2+\f{51}{128}e^4\big)\ep^2_q-\f{e^2}{3}\big(1+\f{e^2}{4}\big)\ep_m\ep_q-\f{1}{15}\big(1-18e^2-\f{39}{8}e^4\big)\ep_d\ep_o\bigg\},
\end{split}
\end{align}
\begin{align}
\begin{split}\label{e_t_derivative_1}
\dot{e}=&-\f{G^2\mu m}{4a^3}\ep_d^2\f{e}{(1-e^2)^{3/2}}-\f{G^3\mu m^2}{a^4}\f{e}{(1-e^2)^{5/2}}\bigg\{\f{1}{15}\big(304+121e^2\big)\big(1+\ep_1\ep_2\big)
\\&
+\f{1}{4}\big(1+\f{e^2}{4}\big)\ep^2_m-\f{1}{6}\Big[7\big(1+\f{e^2}{4}\big)\ep_{d1}+\big(11+\f{13}{2}e^2\big)\ep_{d2}\Big]\ep_d+\f{103}{60}\big(1+\f{163}{412}e^2\big)\ep_q^2
\\&
-\f{1}{6}\big(1+\f{e^2}{4}\big)\ep_m\ep_q+\f{17}{60}\big(1+\f{107}{68}e^2\big)\ep_d\ep_o\bigg\}.
\end{split}
\end{align}
%
From these and the Kepler's third law, we have $\dot{e}/e\sim\dot{a}/a\sim\dot{P}_b/P_b$, where $P_b$ is the orbital period of the binary pulsar system. Because, generally, the measurement of the orbital period decay rate $\dot{P}_b$ is more accurate than that of $\dot{e}$ (or $\dot{a}$), the constraint on the theory from $\dot{P}_b$ is stronger, which was discussed in \cite{Zhang:2017aa}. The semimajor axis and eccentricity decay more rapidly than in GR due to the emission of dipole radiation, unless $\ep_d=\ep_2-\ep_1=0$. All terms leading to deviations away from GR are due to contributions of the scalar field. It can be seen that the above expressions all reduce to the GR results in the limiting case $\ep_1$ and $\ep_2\rightarrow 0$.

%
\begin{figure}[!htbp]
\centering
\includegraphics[width=10cm, height=7.5cm]{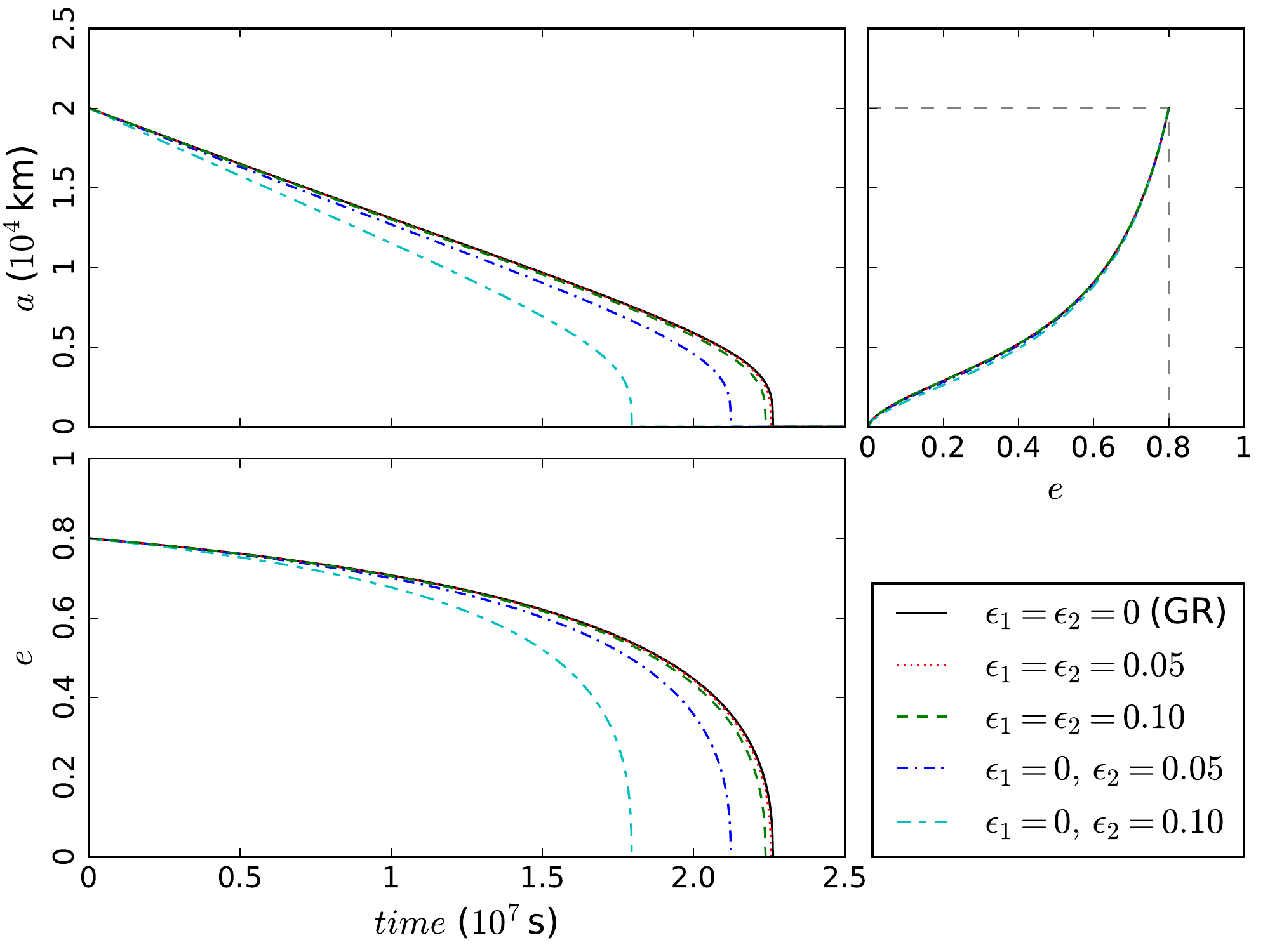}
\caption{Temporal evolution of the semi-major axis ($a$) and the orbital eccentricity ($e$) of a $1.0M_\odot$-$1.6M_\odot$ binary system due to the modified decay rates of the orbital energy and angular momentum. To construct this figure, we used an initial orbital eccentricity of 0.8 and an initial semi-major axis of $2.0\times10^4$ km. Observe that they decay faster than in GR. }
\label{fig_ae}
\end{figure}
%

In Figure~\ref{fig_ae}, we show the temporal evolution of the semi-major axis and the orbital eccentricity for the different combination of the scalar charges of binary system. This evolution was obtained by numerically solving Eqs.~\eqref{a_t_derivative_1} and~\eqref{e_t_derivative_1} for a binary with masses $1.0 M_{\odot}$-$1.6 M_{\odot}$ and an initial orbital eccentricity of 0.8 and an initial semi-major axis of $2.0\times10^4$ km (correspond to an initial angular frequency of 0.2 Hz). We can see that the semi-major axis and the orbital eccentricity decay typically faster in SMG than in GR ($\ep_1=\ep_2=0$), especially when $\ep_2-\ep_1\neq0$ due to the presence of dipole radiation. Such increased decay rates will imprint onto the chirping rate of GWs, which could be constrained by GWs observations in the future.

\section{Conclusion}\label{section5}
SMG is a kind of scalar-tensor theory with screening mechanisms, which can suppress the fifth force in dense regions and allow theories to evade the tight gravitational tests in the solar system and the laboratory. In SMG, all modifications are suppressed by the object's scalar charge, which is inversely proportional to its compactness. Thus, the deviations from GR become small for strongly self-gravitating bodies, which is completely different from other alternative theories without screening mechanisms.

The recent GWs observations provide an excellent opportunity to perform quantitative tests of dissipative sector and strong-field dynamics of gravity theories. The interpretation of GWs observations relies on the comparison of the data to the GWs model, which depends on the orbital evolution of the binary systems.

In this paper, we investigated how the screening mechanisms in SMG affect the orbital evolution of eccentric compact binary systems. From the theory's Lagrangian by investigating the Noether charges and currents, we calculated the angular momentum flux both as a function of the fields in the theory and as a function of the multipole moments of a N-body system. We then specialized to the orbital parameters of  eccentric compact binaries to express its orbital angular momentum decay. This calculation, together with the GWs energy flux of~\cite{Zhang:2017aa}, allowed us to compute the decay rates of the orbital eccentricity and semimajor axis of the binary systems. It turned out that these decay rates are typically faster in SMG than in GR due to the emission of dipole radiation by the scalar mode of the theory. Such modifications can imprint onto the GWs signal and leave a signature of SMG, which could be constrained in the future with GWs observations.

At the end of this paper, we would like to emphasize that the results in this article are applicable to the general SMG, including the chameleon, symmetron, dilaton and $f(R)$ theories.

\begin{acknowledgments}
This work is supported by NSFC No. 11603020, 11633001, 11173021, 11322324, 11653002, 11421303, project of Knowledge Innovation Program of Chinese Academy of Science, the Fundamental Research Funds for the Central Universities and the Strategic Priority Research Program of the Chinese Academy of Sciences Grant No. XDB23010200.
\end{acknowledgments}

\bibliography{}

\end{document}